\documentclass{article}

\usepackage{enumitem}
\usepackage{spconf,amsmath,graphicx}
\usepackage{algorithm}
\usepackage{textcomp}
\usepackage{xcolor}
\usepackage{subcaption}
\usepackage{amsmath}
\usepackage{bm}
\usepackage{caption}

\usepackage{amsmath, amssymb}
\usepackage{amsmath}
\usepackage{amsfonts}
\usepackage{algpseudocode}

\title{Conditional Prior-based Non-stationary Channel Estimation Using Accelerated Diffusion Models}
\name{
\begin{tabular}{c}
Muhammad Ahmed Mohsin\textsuperscript{1}, 
Ahsan Bilal\textsuperscript{2},
Muhammad Umer\textsuperscript{1},
Asad Aali\textsuperscript{3},\\
Muhammad Ali Jamshed\textsuperscript{4},
Dean F. Hougen\textsuperscript{2},
John M. Cioffi\textsuperscript{1}
\end{tabular}
} 
\address{
 \textsuperscript{1} Dept. of Electrical Engineering, Stanford University, Stanford, CA, USA\\
 \textsuperscript{2} School of Computer Science, University of Oklahoma, Norman, OK, USA\\
 \textsuperscript{3} Dept. of Radiology, Stanford University, Stanford, CA, USA\\
 \textsuperscript{4} College of Science and Engineering, University of Glasgow, Glasgow, UK
}

\begin{document}
%
\maketitle
\begin{abstract}
Wireless channels in motion-rich urban microcell (UMi) settings are non-stationary; mobility and scatterer dynamics shift the distribution over time, degrading classical and deep estimators. This work proposes \emph{conditional prior diffusion} for channel estimation, which learns a history-conditioned score to denoise noisy channel snapshots. A temporal encoder with cross-time attention compresses a short observation window into a context vector, which captures the channel's instantaneous coherence and steers the denoiser via feature-wise modulation. In inference, an SNR-matched initialization selects the diffusion step whose marginal aligns with the measured input SNR, and the process follows a shortened, geometrically spaced schedule, preserving the signal-to-noise trajectory with far fewer iterations. Temporal self-conditioning with the previous channel estimate and a training-only smoothness penalty further stabilizes evolution without biasing the test-time estimator. Evaluations on a 3GPP benchmark show lower NMSE across all SNRs than LMMSE, GMM, LSTM, and LDAMP baselines, demonstrating stable performance and strong high SNR fidelity.
\end{abstract}
\begin{keywords}
channel estimation, diffusion models, non-stationary channels, MIMO systems
\end{keywords}
\section{Introduction}\label{sec:intro}
Accurate channel state information (CSI) is a first-order requirement for modern wireless systems. In wideband multiple-input multiple-output (MIMO) deployments, reliable channel estimation is central to realizing spatial multiplexing gains~\cite{liang2014low}. In urban microcells (UMi), however, user acceleration compresses temporal coherence through time-varying Doppler, which degrades pilot-aided interpolation and yields stale CSI~\cite{khalili2015effect, abdolee2013diffusion, umer2025neuralgaussianradiofields, 10888845}. Such \emph{non-stationarity}, characterized by time-varying tap statistics and correlation lengths, breaks stationary assumptions and degrades models trained on fixed distributions.

Recent works have explored sequence models, such as recurrent encoders, for time-varying MIMO channels~\cite{rakhmania2025channel}. However, these approaches implicitly assume slowly varying (piecewise-stationary) statistics and tend to entangle measurement noise with channel dynamics, often requiring frequent re-training as coherence shrinks. In contrast, diffusion models offer a principled alternative by decoupling the signal from noise via a controlled schedule and score matching, enabling calibrated denoising under rapidly changing conditions~\cite{arvinte2022mimo, 11011078}. Yet, standard diffusion-based denoisers are typically \emph{unconditional} and thus ignore the history-dependent structure induced by dynamics such as changing user speeds~\cite{fesl2024diffusion}.

This paper addresses non-stationary channel estimation by learning a \emph{conditional prior} from a causal window of channel observations. The prior is injected into the reverse diffusion process to align the denoising updates with the channel's current coherence length. The temporal context is encoded using a lightweight recurrent encoder with cross-time attention, and a \emph{Gauss-Markov self-conditioning} mechanism feeds the previous clean estimate into the prior to preserve physically plausible evolution. The approach fuses diffusion time, sequence index, and temporal context via Feature-wise Linear Modulation (FiLM) to steer features without architectural bloat. To reduce sampling cost, an \emph{SNR-matched initialization}~\cite{fesl2024asymptotic} starts the reverse process at the diffusion step whose marginal distribution best matches the measured SNR, followed by traversal of a subsampled, geometrically spaced diffusion ladder.

\textbf{Contributions.} The main contributions are as follows.
\begin{enumerate}[leftmargin=*, parsep=0pt]
    \item A conditional diffusion framework for non-stationary channel estimation that learns history-conditioned priors from causal observation windows. The framework uses ConvLSTM-based temporal encoding with cross-time attention and Gauss-Markov self-conditioning to ensure temporal stability.
    \item An accelerated inference procedure that employs SNR-matched initialization and geometrically spaced sampling schedules. This approach reduces computational complexity while preserving signal-to-noise trajectories using significantly fewer iterations.
    \item State-of-the-art performance on 3GPP UMi benchmarks with mobile users at varying speeds ($2\text{-}4$ to $30\text{-}80$ mph). The proposed framework achieves $-17.3$ dB mean NMSE (representing a $2.3$ dB gain over the best baseline) and operates within $2\text{-}4$ dB of oracle performance at high SNR.
\end{enumerate}
\section{Problem Formulation}
\label{sec:prob_form}
A multiuser OFDM downlink in a UMi scenario is considered, with $\mathit{U}$ users, $\mathit{F}$ active subcarriers (tones), and a discrete time (slot) index $k \in \mathcal{K} \triangleq \{0,1,\dots,K-1\}$. Let $\mathcal{U} \triangleq \{1,\dots,U\}$ denote the user set and $\mathcal{F} \triangleq \{1,\dots,F\}$ the tone set. Throughout this work, the parameters are set to $U=8$ and $F=52$\footnote{$U=8$ captures multiuser scheduling scenarios, while $F=52$ corresponds to a standard OFDM resource block size following pilot and guard tone removal~\cite{agheli2025learning}.}. Unless otherwise specified, the system employs $N_t$ transmit and $N_r$ receive antennas, and all notation refers to baseband frequency-domain quantities.

For a given user $u$ at time slot $k$ and tone $f$, the received vector $Y_{u,k}[f]\in\mathbb{C}^{N_r}$ is related to the transmitted symbol vector $X_{u,k}[f]\in\mathbb{C}^{N_t}$ and the MIMO channel frequency response $H_{u,k}[f] \in \mathbb{C}^{N_r \times N_t}$ by the linear model
\begin{equation}
Y_{u,k}[f] = H_{u,k}[f] X_{u,k}[f] + N_{u,k}[f],
\label{eq:per-user}
\end{equation}
where $N_{u,k}[f] \sim \mathcal{CN}(0,\sigma_{u,k}^2 I)$ represents additive white Gaussian noise. In practice, the instantaneous channel $H_{u,k}[f]$ is unknown and must be estimated from pilots transmitted on a known frequency-time subset $\mathcal{P}_u\subseteq \mathcal{K}\times\mathcal{F}$. Application of a coarse per-tone estimator (e.g., LS or MMSE) at these pilot locations produces a noisy channel sample
\begin{equation}
\tilde{H}_{u,k}[f] = H_{u,k}[f] + W_{u,k}[f],
\label{eq:noisy-CSI}
\end{equation}
where $W_{u,k}[f]\sim\mathcal{CN}(0,\tau_{u,k,f}^2 I)$ represents the channel estimation error. Interpolation across the frequency-time grid subsequently yields a pilot-aided \emph{noisy CSI field} $\tilde{H}_k\triangleq\{\tilde{H}_{u,k}[f]\}_{u,f}$ that approximates the true channel field $H_k\triangleq\{H_{u,k}[f]\}_{u,f}$.

The primary challenge arises from user mobility. Each user $u\in\mathcal{U}$ moves with a time-varying speed $v_u(k)$, inducing a corresponding Doppler frequency $f_{D,u}(k) = \tfrac{f_c}{c} v_u(k)$ that causes time selectivity and channel aging. This mobility-induced \emph{non-stationarity} implies that the channel's statistical properties evolve over time. The channel exhibits local correlation, however, and its second-order temporal evolution can be characterized by the time-varying correlation matrix
\begin{equation}
\mathbf{R}^{(\tau)}_k \triangleq \mathbb{E}\!\left[\mathrm{vec}(H_k)\,\mathrm{vec}(H_{k-\tau})^{H}\right],\quad \tau\in\mathbb{Z}.
\label{eq:temporal-corr}
\end{equation}
Here, $\mathbf{R}^{(\tau)}_k$ varies with the time index $k$ through the set of user velocities $\{v_u(k)\}$. Under isotropic scattering conditions, the per-user temporal correlation scales approximately as $\rho_u^{(\tau)}(k)\approx J_0\big(2\pi f_{D,u}(k)\tau\Delta t\big)$, directly relating the channel's correlation length to user mobility~\cite{yang2019deep, rao2017adaptive}.

To address this estimation challenge using a data-driven approach, the problem is reframed as a denoising task on a sequence of channel snapshots. Let $\{x_k\}_{k\in\mathbb{Z}}$ denote the discrete-time sequence of clean channel snapshots\footnote{Each latent, noise-free CSI snapshot $x_k$ (with real and imaginary components stacked) serves as ground truth in simulations and as an oracle-quality reference for real data.}, where each $x_k\in\mathbb{R}^{C\times H\times W}$ represents a real-valued tensor representation of the complex CSI. Here, $C=2$ accommodates the stacked real and imaginary components, $H$ denotes the antenna dimension (e.g., $H = N_t \times N_r$), and $W$ represents the frequency-time grid size. The corresponding noisy observation is modeled as
\begin{equation}
y_k = x_k + n_k, \quad n_k \sim \mathcal{N}(0,\sigma_k^2 I), \quad \mathrm{SNR}_k = \frac{\mathbb{E}\|x_k\|_2^2}{\sigma_k^2}.
\end{equation}
The objective is to recover the clean channel $x_k$ from the noisy observation $y_k$ by exploiting the temporal correlation captured within a causal window $\mathcal{T}_k=\{y_{k-T_w+1},\dots,y_{k-1}\}$, where $T_w\geq 2$ denotes the temporal window length. Since the learning framework operates on real-valued tensors, the complex channel matrices are converted using a real-imaginary stacking operation defined as $x_k = [\Re(H_k); \Im(H_k)]$. To maintain statistical correctness, proper noise scaling is enforced during training. A standard complex Gaussian noise sample is constructed from two real Gaussian components, $\varepsilon_{\Re},\varepsilon_{\Im}\sim\mathcal{N}\left(0,\tfrac{1}{2}I\right)$, such that $\varepsilon = [\varepsilon_{\Re};\varepsilon_{\Im}]$ satisfies $\varepsilon \sim \mathcal{N}(0,I)$ and the complex equivalent exhibits unit variance: $\mathbb{E}[|\varepsilon_{\Re}+j\varepsilon_{\Im}|^2]=1$.
\begin{figure}[t]
\centering
\includegraphics[width=1\columnwidth]{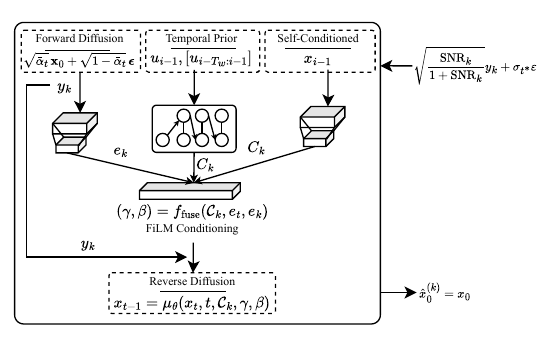}
\caption{Combining noisy forward diffusion, ConvLSTM-attention temporal priors, and self-conditioning to guide reverse diffusion via FiLM parameters.}
\label{fig:diffusion_architecture}
\end{figure}
\begin{figure*}[t!]
  \centering
  \begin{subfigure}[t]{0.66\columnwidth}
    \centering
    \includegraphics[width=\textwidth]{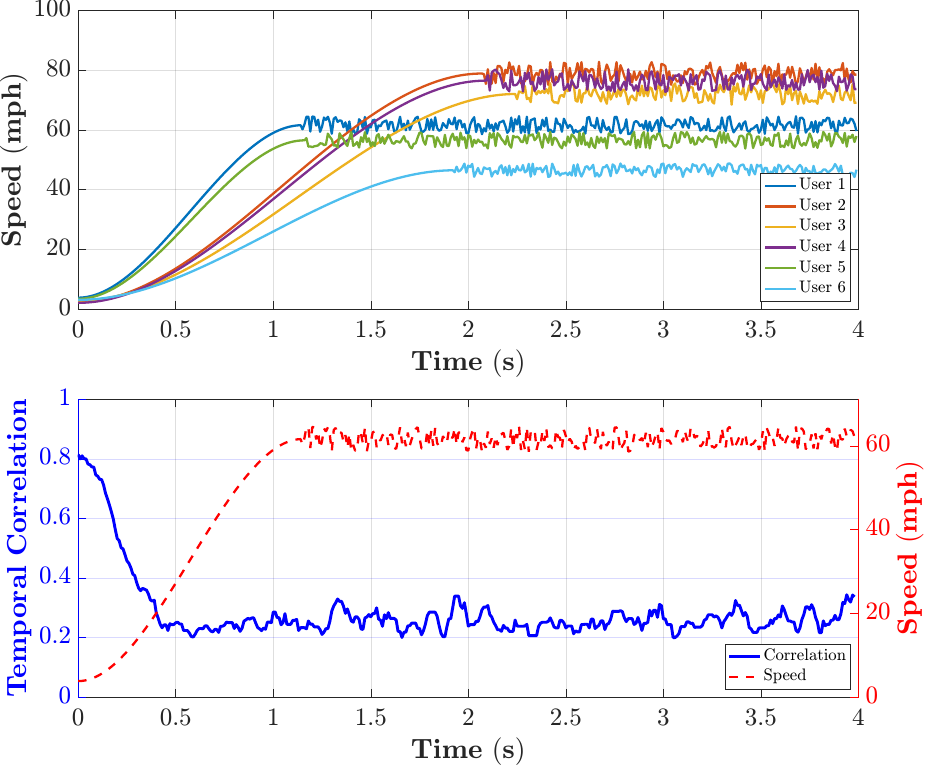}
    \caption{User speed and correlation decay in UMi}
    \label{fig:EpochsvLoss}
  \end{subfigure}
  \hspace{-0.01\columnwidth}
  \begin{subfigure}[t]{0.66\columnwidth}
    \centering
    \includegraphics[width=\textwidth]{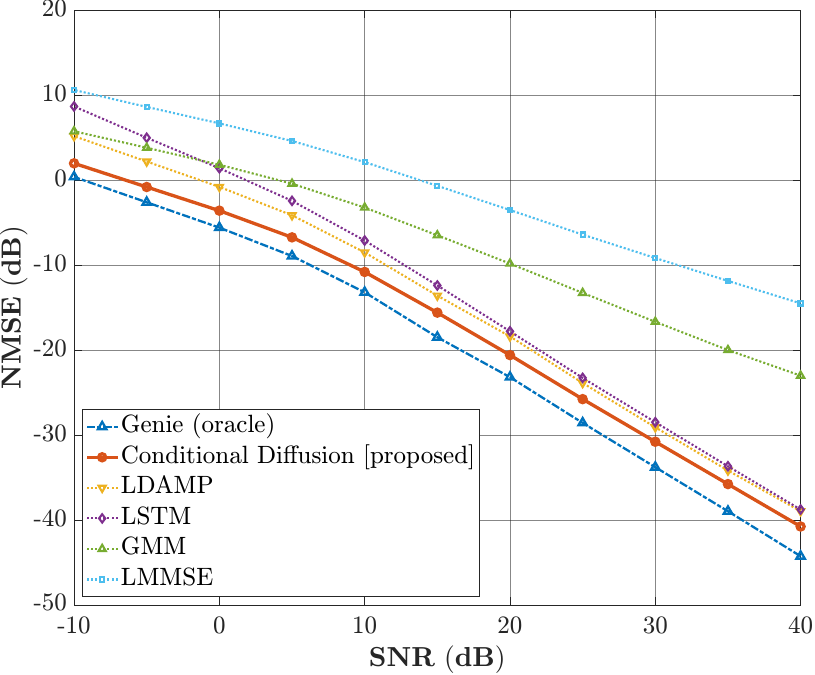}
    \caption{NMSE vs.\ SNR: diffusion vs.\ baselines}
    \label{fig:EpochsvNMSE}
  \end{subfigure}
  \hspace{-0.01\columnwidth}
  \begin{subfigure}[t]{0.66\columnwidth}
    \centering
    \includegraphics[width=\textwidth]{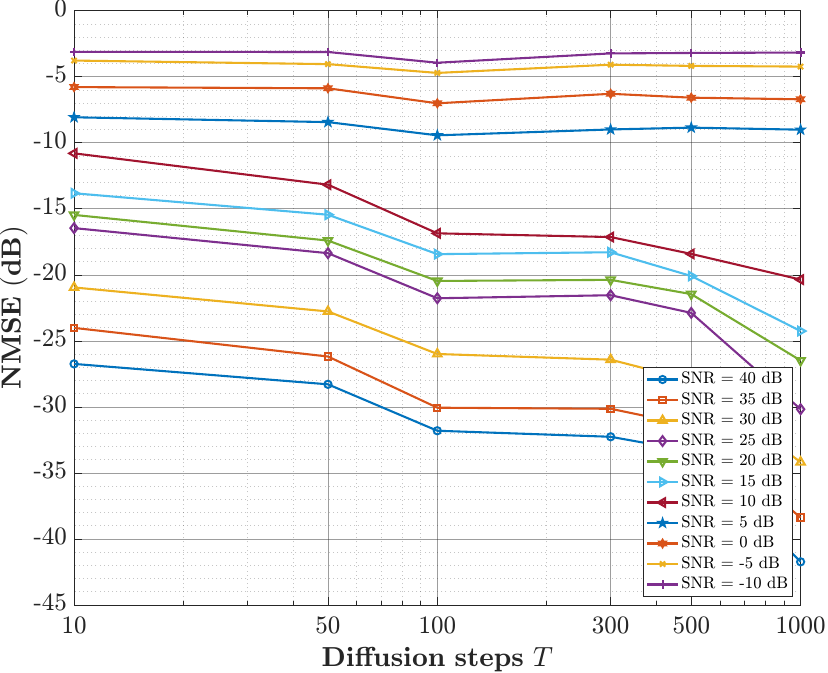}
    \caption{NMSE vs.\ diffusion steps across SNRs}
    \label{fig:SNRvNMSE}
  \end{subfigure}
  \caption{UMi benchmark and diffusion results: \textbf{(a)} mobility profile and correlation, \textbf{(b)} accuracy across SNRs, \textbf{(c)} convergence vs.\ diffusion steps.}
  \label{fig:2}
\end{figure*}

\section{Conditional Prior Based Non-Stationary Estimation}
\subsection{Forward Diffusion Process}\label{subsec:forward}
To enable principled denoising through score matching, a forward diffusion process is employed that gradually corrupts the clean channel snapshot with Gaussian noise. Given a monotonic noise schedule $\beta_t \in (0,1)$ for $t=1,\ldots,T$ diffusion steps, the cumulative noise parameters are defined as $\alpha_t = 1-\beta_t$ and $\bar{\alpha}_t = \prod_{s=1}^{t}\alpha_s$. The forward process admits the following closed-form solution, allowing direct sampling at any diffusion step $t$
\begin{align}
q(x_t \mid x_0) &= \mathcal{N}(\sqrt{\bar{\alpha}_t}\,x_0,\; (1-\bar{\alpha}_t)I), \label{eq:forward_marginal}\\
x_t &= \sqrt{\bar{\alpha}_t}\,x_0 + \sqrt{1-\bar{\alpha}_t}\,\varepsilon,\quad \varepsilon \sim \mathcal{N}(0,I). \label{eq:forward_sample}
\end{align}
The noise variance at step $t$ is $\sigma_t^2 = 1-\bar{\alpha}_t$, which increases monotonically from $\sigma_1^2 \approx 0$ to $\sigma_T^2 \approx 1$. This noise schedule defines a per-step signal-to-noise ratio (SNR), $\mathrm{SNR}(t)=\bar{\alpha}_t/(1-\bar{\alpha}_t)$, which is subsequently used to enable SNR-matched reverse initialization (Sec.~\ref{subsec:reverse}).

\subsection{Conditional Prior and Training Objective}\label{subsec:cond-prior}
The diffusion process is conditioned on temporal context to enable adaptation to time-varying channel statistics. The history $\mathcal{T}_k$ is compressed into a context vector $\mathcal{C}_k \in \mathbb{R}^{d_c}$, which defines a conditional prior $p(x_0 \mid \mathcal{C}_k)$ over the clean snapshot. Let $p_t(x_t \mid \mathcal{C}_k)$ denote the $\mathcal{C}_k$-conditional marginal at step $t$. The reverse diffusion is then guided by the conditional score function
\begin{equation}
s^\star(x_t,t\mid \mathcal{C}_k) = \nabla_{x_t}\log p_t(x_t\mid \mathcal{C}_k). \label{eq:conditional_score}
\end{equation}
Using the noise parameterization and the score–noise identity, this score is equivalent to
\begin{equation}
s^\star(x_t,t\mid \mathcal{C}_k) = -\frac{1}{\sigma_t}\,\mathbb{E}[\varepsilon \mid x_t,\mathcal{C}_k], \label{eq:score_noise_identity}
\end{equation}
where $\varepsilon = (x_t-\sqrt{\bar{\alpha}_t}x_0)/\sigma_t$. A neural denoiser $\varepsilon_\theta$ is trained to approximate the conditional expectation of the noise by sampling $t\sim\mathrm{Unif}\{1,\ldots,T\}$ and $\varepsilon\sim\mathcal{N}(0,I)$ to minimize the objective
\begin{equation}
\mathcal{L}_{\text{noise}}(\theta) = \mathbb{E}_{x_0,t,\varepsilon}\!\left[\|\varepsilon - \varepsilon_\theta(x_t,t,\mathcal{C}_k)\|_2^2\right]. \label{eq:noise_loss}
\end{equation}
This objective corresponds to conditional score matching, where the learned denoiser provides an estimate of the score function: $\hat{s}_\theta(x_t,t \mid \mathcal{C}_k) = -\frac{1}{\sigma_t}\varepsilon_\theta(x_t,t,\mathcal{C}_k)$.

\subsection{Temporal Context Encoding}\label{subsec:temp-encoder}
The effectiveness of the conditional approach depends critically on the construction of $\mathcal{C}_k$ from the history $\mathcal{T}_k$. A three-stage pipeline is used to capture both spatial structure and temporal dependencies:

\emph{Stage 1: Spatial Encoding.} A shared CNN $\phi_{\text{pre}}$ with time-independent parameters processes each observation to extract spatial features
\begin{equation}
z_i = \phi_{\text{pre}}(y_i) \in \mathbb{R}^{d\times H'\times W'},\quad i \in \{k-T_w+1,\dots,k-1\}. \label{eq:spatial_encoding}
\end{equation}

\emph{Stage 2: ConvLSTM Aggregation.} A ConvLSTM updates its hidden and cell states while preserving the spatial grid of the features
\begin{equation}
(h_i, c_i) = \mathrm{ConvLSTM}(z_i, h_{i-1}, c_{i-1}). \label{eq:convlstm}
\end{equation}
Each hidden state $h_i$ is spatially pooled to produce a vector $u_i$, and these vectors are stacked into a sequence matrix $U=[u_{k-T_w+1},\dots,u_{k-1}] \in \mathbb{R}^{(T_w-1)\times d}$.

\emph{Stage 3: Cross-Time Attention.} Using the most recent state $q=u_{k-1}$ as the query and the historical sequence $U$ as keys and values, a cross-attention mechanism produces the final temporal context vector $\mathcal{C}_k$. This setup yields a compact representation that tracks the instantaneous coherence length, which the reverse process subsequently exploits.

\subsection{FiLM Conditioning}\label{subsec:film}
Given the context $\mathcal{C}_k$, the denoiser $\varepsilon_\theta$ is conditioned using three embeddings: the diffusion step $e_t = f_t(\mathrm{PE}(t))$, the sequence index $e_k = f_k(\mathrm{PE}(k))$, and the temporal context $e_{\text{temp}} = f_{\text{temp}}(\mathcal{C}_k)$. These embeddings are fused to generate FiLM parameters
\begin{equation}
z_k = f_{\text{fuse}}([e_t; e_k; e_{\text{temp}}]) \in \mathbb{R}^{2d}, \label{eq:fuse}
\end{equation}
which split as $z_k=(\gamma_k,\beta_k)$ with $\gamma_k,\beta_k\in\mathbb{R}^d$. Feature-wise linear modulation gates spatial features via $\tilde{v}=(1+\gamma_k)\odot\phi_{\text{pre}}(x_t)+\beta_k$, and $\phi_{\text{post}}$ decodes $\tilde{v}$ to the noise prediction. FiLM couples diffusion time, temporal context, and sequence position without architectural complexity, steering denoising along context-relevant directions.

\subsection{Reverse Process with Acceleration}\label{subsec:reverse}
During inference, the forward process is reversed using the trained denoiser. The reparameterized mean of the reverse step is
\begin{equation}
\mu_\theta(x_t,t,\mathcal{C}_k) = \frac{1}{\sqrt{\alpha_t}}\!\left(x_t - \frac{\beta_t}{\sqrt{1-\bar{\alpha}_t}}\varepsilon_\theta(x_t,t,\mathcal{C}_k)\right). \label{eq:reverse_mean}
\end{equation}
To accelerate inference, the process is initialized at the diffusion step whose marginal SNR matches the measured input SNR. The starting step $t^*$ is selected as
\[
t^* \!=\! \arg\min_{t\in\{1,\dots,T\}}\! \big|\mathrm{SNR}(t)-\mathrm{SNR}_k\big|,
\]
and the initial state is set to $x_{t^*}^{(0)} = \sqrt{\tfrac{\mathrm{SNR}_k}{1+\mathrm{SNR}_k}}\,y_k$. The process then traverses a shortened schedule $\mathcal{S}=\{t_1=t^* > t_2 > \cdots > t_L=0\}$ with geometric spacing $t_\ell=\big\lfloor r^{\,\ell-1} t^* \big\rfloor$ for some ratio $r \in (0,1)$. This approach preserves the signal-to-noise trajectory while using far fewer steps ($L \ll T$).

\subsection{Temporal Self-Conditioning}\label{subsec:self-cond}
To utilize inter-slot consistency, the previous clean estimate $\hat{x}_0^{(k-1)}$ is injected into the context vector $\mathcal{C}_k$. This creates an implicit Gauss-Markov prior that reduces estimation variance via the law of total variance
\begin{equation}
\mathrm{Var}(x_0^{(k)}\mid \mathcal{I}_k,\hat{x}_0^{(k-1)}) \leq \mathrm{Var}(x_0^{(k)}\mid \mathcal{I}_k), \label{eq:variance_reduction}
\end{equation}
where $\mathcal{I}_k$ represents all available information at time $k$. The clean prediction at any step $t$ is given by
\begin{equation}
\hat{x}_0(x_t,t) = \frac{1}{\sqrt{\bar{\alpha}_t}}\left(x_t-\sqrt{1-\bar{\alpha}_t}\,\varepsilon_\theta(x_t,t,\mathcal{C}_k)\right). \label{eq:clean_prediction}
\end{equation}
During training, gradients are stopped through $\hat{x}_0^{(k-1)}$ (i.e., using $\mathrm{sg}[\hat{x}_0^{(k-1)}]$) to prevent feedback instabilities while still enabling the model to learn the appropriate temporal coupling.

\subsection{Temporal Regularization}\label{subsec:temp-reg}
To complement self-conditioning and ensure physically plausible evolution, a temporal smoothness term is added during training only
\begin{equation}
\mathcal{L}_{\text{total}} = \mathcal{L}_{\text{noise}} + \lambda\|\hat{x}_0^{(k)} - \hat{x}_0^{(k-1)}\|_2^2, \qquad \lambda \ll 1. \label{eq:total_loss}
\end{equation}
This regularizer encourages successive estimates to be consistent with the mobility-implied coherence while leaving the test-time estimator unbiased.

\section{Experimental Results \& Analysis}
The conditional diffusion framework is benchmarked in a simulated 3GPP UMI environment at a carrier frequency of $f_c=3.5$ GHz with $80$ MHz bandwidth. This benchmark features mobility-induced non-stationarity, where user speeds vary over time, starting at $v_u(0)\in[2,4]$ mph and accelerating to $v_u(K-1)\in[30,80]$ mph over a period of $t_{\mathrm{accel}}\in[1.0,2.5]$ s. This acceleration increases Doppler spread, shortens coherence time, and accelerates channel aging (Fig.~\ref{fig:EpochsvLoss}). The faster decay of correlation under higher speeds validates the problem setting and highlights the need for temporally aware estimation.
 
The proposed framework couples temporal priors with diffusion-based denoising to adapt to non-stationary channels. It consistently outperforms LMMSE~\cite{arellano2024performance}, GMM~\cite{ali2025gaussian}, LSTM~\cite{ranjan2025deep}, and LDAMP~\cite{he2018deep}, approaching near-oracle performance at high SNR while remaining robust in low-SNR regimes (Fig.~\ref{fig:EpochsvNMSE}). These results demonstrate the effectiveness of integrating temporal priors with diffusion-based denoising for non-stationary channels.

The computational efficiency analysis in Fig.~\ref{fig:SNRvNMSE} reveals that the framework adapts computational effort to task complexity through SNR-dependent convergence behavior. High SNR conditions achieve rapid convergence, requiring only a fraction of the total iterations, while low SNR scenarios benefit from extended denoising schedules to fully exploit the diffusion process. This adaptive trade-off demonstrates that the proposed model maintains estimation quality without unnecessary computational overhead across diverse conditions.

\section{Conclusion}
This paper introduced a conditional prior-based diffusion model for non-stationary channel estimation. The proposed framework, which incorporates a ConvLSTM-attention encoder and FiLM conditioning, achieves near-oracle accuracy with reduced sampling complexity. Future extensions of this work include scaling the framework to massive MIMO systems, adapting it to diverse propagation environments exhibiting stronger non-stationarity, and integrating federated learning for distributed deployments.

\bibliographystyle{IEEEbib}
\bibliography{main}

\end{document}